\documentclass[twocolumn]{revtex4}
\usepackage{graphicx}
\usepackage{amsmath}
\usepackage{epsfig}
\usepackage{amssymb}
\usepackage{amsfonts}

\newtheorem{theorem}{Theorem}[section]

\newcommand{\qed}{\nobreak \ifvmode \relax \else
      \ifdim\lastskip<1.5em \hskip-\lastskip
      \hskip1.5em plus0em minus0.5em \fi \nobreak
      \vrule height0.75em width0.5em depth0.25em\fi}
\makeatletter
\def\btt#1{\texttt{\@backslashchar#1}}
\DeclareRobustCommand\bblash{\btt{\@backslashchar}}
\makeatother
\newcommand{\beq}{\begin{equation}}
\newcommand{\eeq}{\end{equation}}

\def\notpa{\hbox{{$\partial$}\kern-.54em\hbox{\raisenot}}}
\def\notA{\hbox{{$A$}\kern-.54em\hbox{\raisenot}}}

\begin{document}

\title{Universal Leakage Elimination}

\author{Mark S. Byrd$^1$}
\author{Daniel A. Lidar$^2$}
\author{Lian-Ao Wu$^2$}
\author{Paolo Zanardi$^3$}
\affiliation{1. Physics Department, Southern Illinois University, Carbondale, Illinois 62901-4401, \\ 
2. Chemical Physics Theory Group, Chemistry Department, and Center for Quantum Information and Quantum Control, University of Toronto, 80 St. George St., Toronto, Ontario M5S 3H6, Canada \\
3.  Institute for Scientific Interchange (ISI), Villa Gualino, Viale Settimio
Severo 65, I-10133 Torino, Italy}

\begin{abstract}
``Leakage'' errors are particularly serious errors which couple states
within a code subspace to states outside of that subspace thus destroying
the error protection benefit afforded by an encoded state. We generalize an
earlier method for producing leakage elimination decoupling operations and
examine the effects of the leakage eliminating operations on
decoherence-free or noiseless subsystems which encode one logical, or
protected qubit into three or four qubits. We find that by eliminating the
large class of leakage errors, under some circumstances, we can create the
conditions for a decoherence free evolution. In other cases we identify a
combination decoherence-free and quantum error correcting code which could
eliminate errors in solid-state qubits with anisotropic exchange interaction
Hamiltonians and enable universal quantum computing with only these
interactions.
\end{abstract}

\maketitle

\pagebreak


\section{INTRODUCTION}

Noise protection for quantum information processing is an important facet of
quantum control and the design of quantum devices. In quantum computing,
coherent control of a quantum system is required in order to take advantage
of quantum computing speed-ups. A great deal of work has been done, and is
still ongoing, to try to achieve multi-particle control for quantum
information processing. In order to implement noiseless control of quantum
computing systems, several methods of error prevention have been introduced.
Quantum error correcting codes (QECCs) \cite%
{Shor:95,Calderbank:96,Steane,Gottesman:97,Gottesman:97b,Knill:97b}, detect
and correct errors, decoherence-free or noiseless, subsystems (DFSs) \cite%
{Zanardi:97c,Duan:98,Lidar:PRL98,Knill:99a,Kempe:00,Lidar:00a} avoid noises
in quantum systems and dynamical decoupling controls (DD) \cite%
{Viola:98,Ban,Duan:98e,Viola:99,Zanardi:98b,Vitali:99,Byrd/Lidar:ebb,Viola:99a,Viola:00a,Vitali:01,Byrd/Lidar:01,Cory:00,Agarwal:01,Uch:02,Uch:03}
reduce the errors by averaging or symmetrizing them away. Since none of
these has seen the ultimate success of preventing errors in a prototypical
quantum computing device, combinations of more than one of these methods
have been explored \cite%
{Lidar:PRL99,Zanardi:98b,Zanardi:99a,Viola:01a,Byrd/Lidar:ss,Byrd/Lidar:pqe02,Alber/etal:01,Wu/Lidar:cdfs,Khodjasteh/Lidar:02,Lidar/Wu:02,Wu/etal:02}%
. One particularly promising example is the combination of dynamical
decoupling controls with decoherence-free subsystems \cite%
{Zanardi:98b,Viola:00a,Wu/Lidar:cdfs,Byrd/Lidar:ss,Viola:01a,Byrd/Lidar:ebb,Byrd/Lidar:pqe02,Lidar/Wu:02,Wu/etal:02,Zanardi:99a,YZhang/etal:04}%
. This combination can offer several advantages; it can 1) reduce the number
of physical qubits required to encode one logical qubit, 2) can enable
universal control in systems which cannot be completely controlled
otherwise, 3) can avoid noises, 4) can reduce noises even if they are not
eliminated or avoided. Such combinations are very likely to be necessary for
the near-term and longer term goals associated with reliable quantum
information processing. (For a recent review on error prevention, see \cite%
{Byrd/etal:pqe04}).

For those quantum computing proposals which use quantum dots for storing
information and the Heisenberg exchange interaction for performing gating
interactions, a DFS encoding is promising since it enables universal
computing without the need for single qubit gates \cite%
{Bacon:99a,Kempe:00,Bacon:Sydney,DiVincenzo:00a,Levy:01,Benjamin:01,Wu/Lidar:01a,Lidar/Wu:01,Kempe:01d,Kempe:01e,Lidar/Wu/Blais:02}%
. Architecturally, as well as for speed, single qubit gates can be difficult
to implement for unencoded (i.e., physical) qubits in solid-state systems.
It has been shown that, for several different types of interactions, and for
several different encodings, decoherence-free subspaces provide the ability
to perform universal quantum computing without requiring single qubit gates 
\cite%
{Bacon:99a,Kempe:00,Bacon:Sydney,DiVincenzo:00a,Levy:01,Benjamin:01,Wu/Lidar:01a,Lidar/Wu:01,Kempe:01d,Kempe:01e,Lidar/Wu/Blais:02}%
. It is therefore important to know the conditions for a DFS to exists or,
as discussed in this paper, what methods might be used to create a DFS. For
those circumstances which do not allow for a DFS implementation alone in
order to eliminate all noise in the system, what is (are) the best method(s)
for error protection? This clearly depends on the physical system and an
analysis of the types of occuring errors will be necessary in order to take
advantage of every possible technique for noise suppression, correction
and/or avoidance.

In this paper we discuss the elimination of leakage errors. Leakage errors
destroy a subspace encoding by coupling states within the encoded subspace
of the system Hilbert space with the states which are outside of the code
subspace. These are particularly serious errors since they eliminate the
usefulness of a subspace encoding. Moreover, they cannot be handled by
standard QECC methods under the assumption of a set of operations restricted
to act on a subspace \cite{Mohseni/Lidar:inprep}. We will first review the
bang-bang limit of the dynamical decoupling controls and the algebraic
decomposition of the operators on the Hilbert space in Section \ref{sec:rev}
in order to make the article more self-contained. We then review, in Section %
\ref{sec:LEOs}, the definition and construction of leakage elimination
operators (LEOs) using canonical gates and then generalize the construction
to gates which are not canonical. In Section \ref{sec:3qdfs} we provide an
explicit decomposition of the algebra of operators for the 3-qubit DFS and
use it to classify all errors on the DFS. In this section, we also analyze
the errors which commonly arise in solid-state implementations of quantum
computing proposals in terms of the basis set we have constructed and
determine a strategy for eliminating all errors, in addition to leakage
errors, which arise from anisotropic exchange errors and cause decoherence.
In the following section, Section \ref{sec:4qdfs}, we construct a physically
available LEO which is not made of canonical gates using the construction in
Section \ref{sec:LEOs}. We then analyze errors in the 4-qubit DFS which
arise in solid-state implementations of QC. We then summarize our results in
the Conclusion.


\section{UNIVERSAL LEAKAGE ELIMINATION}

\label{sec:rev}

In this section we briefly review dynamical decoupling controls,
symmetrization and the results of \cite{Wu/etal:02}. For a more detailed
discussion of dynamical decoupling controls, the reader is referred to \cite%
{Viola:04}, for the group-theoretical underpinnings see \cite{Zanardi:99a},
for an empirical approach see \cite{Byrd/Lidar:ss,Byrd/Lidar:ebb}, for a
geometrical approach see \cite{Byrd/Lidar:01} or 
\cite{Byrd/etal:pqe04,Viola:04}
for recent reviews. We then provide a general formula for producing a
leakage elimination operator (LEO).


\subsection{Dynamical Decoupling Controls}

\label{sec:dd}

Dynamical decoupling controls are control pulses which are used to average
away noises in a quantum system. When hard, fast pulses are used, these are
commonly referred to as bang-bang controls. Here we review decoupling
controls in the bang-bang limit.

Consider a general Hamiltonian of the form 
\begin{equation*}
H=H_{S}+H_{B}+H_{SB},
\end{equation*}%
where $H_{S}$ acts only on the system, $H_{B}$ acts only on the bath, and $%
H_{I}=H_{SB}=\sum_{\gamma }S_{\gamma }\otimes B_{\gamma }$ couples the
system to the bath. Let us now implement control operations, $U_{i}$,
periodically with the system undergoing free unitary evolution (by $H$) for
a time $\Delta t$ between control operations. If we assume that the free
evolution is negligible during the time the control is \textquotedblleft
on\textquotedblright\ (this assumes \textquotedblleft
strong\textquotedblright\ control Hamiltonians are avaliable.), then we
obtain an effective unitary evolution for the combined system-bath: 
\begin{equation}
U_{eff}\approx \prod_{i=0}^{N-1}U_{i}\exp [-iH\Delta t]U_{i}^{\dagger }.
\end{equation}%
If we also assume that $H$ is approximately constant during the application
of the set of pulses $\{U_{i}\}$ (This assumes that we can apply the pulses
quickly on the system-bath interaction time-scale.), then we may also use an
effective Hamiltonian to describe this evolution, 
\begin{equation}
H_{eff}\approx \frac{1}{N}\sum_{i=0}^{N-1}U_{i}HU_{i}^{\dagger }.
\end{equation}%
In a ideal circumstances (as $N\rightarrow \infty $), we can eliminate $%
H_{SB}$ completely so as to decouple the system and bath. However, \textit{%
in this paper we combine decoupling operations with an encoding, therefore
we only require that }$H_{SB}$\textit{\ be} modified. \textit{This
drastically reduces the demands on a physical system. }

One should note that \textquotedblleft strong\textquotedblright\ and
\textquotedblleft fast\textquotedblright\ are relative to system-bath
interactions, notions which have been thoroughly quantified in \cite%
{Facchi/etal:04b}. In addition, we need not require strong pulses in some
cases \cite{Viola:02,Viola:04} and in other cases, the fast requirement can
be relatively easily satisfied \cite{Shiokawa/Lidar:02,Faoro:03}. In this
paper, we will consider dynamical decoupling controls which assume hard,
fast pulses, but we note that appropriate controls may be available which
can serve as decoupling pulses without the necessity of the ``bang-bang''
limit.

As a final remark on decoupling operations, we state the following theorem 
\cite{Byrd/Lidar:ss,Byrd/Lidar:pqe02} which follows from \cite{Zanardi:99a}:

\begin{theorem}
\label{th1} \textit{Dynamical decoupling with respect to the set of logical
operations of an encoded qubit can be used to completely decouple the
dynamics of the encoded subspace from the bath. }
\end{theorem}

This theorem is important for the following reasons. First, the number of
pulses required to eliminate noise on physical qubits can be quite taxing on
physical resources. If we restrict ourselves to logical operations, we can
reduce the number of required pulses dramatically. Second, in many cases, if
we use logical operations to remove errors, we are restricting to those
operations which are available in experiment. Often an ecoding is chosen for
its universality considerations. In other words, many codes are chosen so
that universal quantum computing can be performed on a subspace even if it
cannot be performed on the entire Hilbert space. Those operations which
achieve universal control, can also be used for complete decoupling.


\subsection{Algebraic Decomposition}

\label{sec:alg}

In order to discuss the effects of the dynamical decoupling operations on
encoded qubits, we will briefly review the decomposition of the algebra \cite%
{Knill:99a,Zanardi:99} which can describe all error prevention schemes \cite%
{Zanardi/Lloyd:03}.

The interaction algebra, denoted $\mathcal{A}$, is generated by the set $%
\{H_{S},S_{\gamma }\}$. This algebra is, in general, reducible and can be
closed under Hermitian conjugation (meaning $\mathcal{A}^{\dagger }=\mathcal{%
A}$). This algebra is a subalgebra of the full set of endomorphisms of the
total Hilbert space $\mathcal{H}$, End($\mathcal{H}$) which are linear
operators on $\mathcal{H}$. The irreducible components of this algebra are
described by the decomposition 
\begin{equation}
\mathcal{A}=\underset{J\in \mathcal{J}}{\oplus }{I}_{n_{J}}\otimes M(d_{J},%
\mathbb{C}),
\end{equation}%
where the $J$, a shorthand for all relevant representation indices, label
the irreducible representations and the $M(d_{J},\mathbb{C})$ are $%
d_{J}\times d_{J}$ complex matrices. This representation is a direct sum
decomposition (block diagonal) with $n_{J}$ labelling the states of the
system in the corresponding Hilbert space decomposition 
\begin{equation}
\mathcal{H}\cong \underset{J\in \mathcal{J}}{\oplus }\mathbb{C}%
^{n_{J}}\otimes \mathbb{C}_{J}^{d}.
\end{equation}%
Each factor $\mathbb{C}^{n_{J}}$ corresponds to a noiseless subsystem. The
commutant of $\mathcal{A}$, denoted $\mathcal{A}^{\prime }$, is in End($%
\mathcal{H}$) and is defined as 
\begin{equation}
\mathcal{A}^{\prime }=\{X\in \mathrm{End}(\mathcal{H})\;|\;[X,\mathcal{A}%
]=0\}.
\end{equation}%
The existence of a decoherence-free, or noiseless, subsystem is equivalent
to 
\begin{equation}
\mathcal{A}^{\prime }=\underset{J\in \mathcal{J}}{\oplus }M(n_{J},\mathbb{C}%
)\otimes {I}_{d_{J}}\neq \mathbb{C}{I}.
\end{equation}%
This implies a non-trivial group of symmetries of the commutant. The unitary
part of $\mathcal{A}^{\prime }$, $U(\mathcal{A}^{\prime })$, is the set of
unitary symmetries of the error algebra $\mathcal{A}$.

Note that DFSs, QECCs and topological codes can all be described by this
same algebraic decomposition \cite{Zanardi/Lloyd:03}. We can therefore
generically discuss encoded qubits in the context of quantum error
prevention without regard to the type of encoding although we will primarily
direct our attention to DFSs.


\section{LEAKAGE AND LEAKAGE ELIMINATION OPERATORS (LEOs)}

\label{sec:LEOs}

Qubits can be either a subspace of a larger system Hilbert space or an
encoded subspace of a larger Hilbert space. The idealized, isolated
two-level system never occurs in nature, when all energy scales are taken
into account. We therefore seek to eliminate, or reduce the difference
between an idealized qubit and the approximate two-level systems available
in experiments. Whether these are two physical states in a larger Hilbert
space, or a state which is encoded into some set of states through a
non-trivial transformation, we will discuss a generalized notion of a code
and encoded subspace. This encoded subspace, or codespace, will be denoted $%
\mathcal{C}$. The othogonal complement of the codespace, also a subspace of
the system Hilbert space, will be denoted $\mathcal{C}^{\perp }$. Our
objective will be to eliminate the coupling between $\mathcal{C}$, and $%
\mathcal{C}^{\perp }$. We refer to such errors as leakage errors. However,
unlike considerations of leakage introduced during logical, or gating
operations \cite%
{Tian:00,Gottesman:97b,Preskill:99,Kempe:01d,Palao:03,Palao:03e,Sklarz/Tannor}, we
will consider residual errors and errors introduce by system-bath couplings
as in \cite{Wu/etal:02}.

Let us first consider the simple case of a physical or encoded qubit. In
this case we would like to eliminate the leakage from a two-level system
within an $N$-level system. We will choose an ordered basis for the $N$%
-level system Hilbert space $\{\left\vert j\right\rangle\}_{j=0}^{N-1}$ such
that the code $\mathcal{C}$ will be represented by some combination of the
first $2$-levels. The algebra of operations on the system Hilbert space can
then be classified in the following way, 
\begin{equation}  \label{eq:EEL}
E = \left(%
\begin{array}{cc}
B & 0 \\ 
0 & 0%
\end{array}%
\right), E^\perp = \left(%
\begin{array}{cc}
0 & 0 \\ 
0 & C%
\end{array}%
\right), L = \left(%
\begin{array}{cc}
0 & D \\ 
F & 0%
\end{array}%
\right),
\end{equation}
where $B$ and $C$ are $2\times 2$ and $(N-2)\times (N-2)$ blocks
respectively, and $D, F$ are $2\times (N-2)$ and $(N-2)\times 2$ blocks
respectively. Operators of the type $E$ represent logical operations, i.e.,
they act entirely within the code subspace. $E^\perp$ operations act only on 
$\mathcal{C}^\perp$ and thus have no effect on the qubit subspace \cite%
{MLEOs:comment1}. Finally, $L$ represents the leakage operators. This
decomposition, for physical or encoded qubits is quite general and the
operators $B$ act only on the logical qubit labels.

Generally, modifying the Hamiltonian (more specifically, the system-bath
interaction Hamiltonian) through the use of dynamical decoupling controls
can change the conditions under which quantum information is protected
against errors. Clearly, this is accomplished by modifying the set $%
\{H_{S},S_{\gamma }\}$, which modifies the interaction algebra $\mathcal{A}$
and thus the irreducible components of $\mathcal{A}$. For example, we can
create DFSs where none were possible without such a modification by inducing
a symmetry \cite%
{Zanardi:98b,Viola:00a,Wu/Lidar:cdfs,Byrd/Lidar:ss,Viola:01a,Byrd/Lidar:ebb,Byrd/Lidar:pqe02,Lidar/Wu:02,Wu/etal:02,Zanardi:99a,YZhang/etal:04}%
. We could also eliminate correlated errors thus changing the requirements
for a QECC. As described in Section \ref{sec:dd}, dynamical decoupling
controls can be seen as a projection onto a subspace of the space of
operators acting on the system Hilbert space.

More to the point of this paper, we can eliminate some of the components of
the interaction algebra by eliminating some of the $S_{\gamma }$ (or some
combination thereof). In our case, we seek to eliminate leakage. We
therefore give a general classification of elements of the algebra according
to their effect on the code space. The elements of the interaction algebra $%
\{A_{i}\}=\mathcal{A}$, will be classified as $\{E_{i}\}\subset \mathcal{A}$%
, which act on the code, $\{E_{i}^{\perp }\}\subset \mathcal{A}$ which
effect the orthogonal subspace, and leakage errors $\{L_{i}\}\subset 
\mathcal{A}$ which couple the elements of the code $\mathcal{C}$ with states
in the orthogonal subspace $\mathcal{C}^{\perp }$.

>From the general decomposition of the algebra in Section \ref{sec:alg}, we
should note the following facts. One can always decompose the algebra in
terms of a set $\mathcal{O}$ which acts on the system Hilbert space and a
set $\mathcal{E}$ which acts on the Hilbert space of the environment. These
sets can be taken to be Hermitian operators with complex coefficients. The
set of operators may then always be expressed as some linear combination of
tensor products of the two sets with complex coefficients. This implies
that, although the symmetries appropriate for a DFS may not always exist, we
may always use a DFS-compatible basis. This will be important in Sections %
\ref{sec:3qdfs} and \ref{sec:4qdfs} where we will discuss a basis for which
a code can be constructed which will protect against errors, even when no
symmetry in the operator algebra exists initially.


\subsection{Canonical LEOs}

In the simplest case of a \textquotedblleft parity-kick\textquotedblright\
bang-bang control \cite{Viola:98,Vitali:99,Zanardi:99}, the decoupling
sequence produces the effective evolution: 
\begin{equation}
H_{eff}\approx \frac{1}{2}\sum_{i=0}^{1}U_{i}HU_{i}^{\dagger },
\label{eq:pkick}
\end{equation}%
i.e., there is only one non-trivial decoupling pulse ($U_{0}\equiv {I}$) 
\cite{BBnote}. We will restrict out attention to parity-kick pulses due to
time constraints which restricts the number of pulses that can be applied in
many physical systems.

Abtractly, we can state the consequences of the parity-kick pulse sequence
as follows. Given any subspace $\mathcal{C}\subset \mathcal{H}$, there is a
canonically associated $\mathbb{Z}_{2}$ group (the cyclic group of order
two). This group is generated by the operator $R_{L}:=\exp (i\pi \Pi _{%
\mathcal{C}})$, where $\Pi _{\mathcal{C}}$ is the projector onto the code
space $\mathcal{C}$. In the language of $\mathbb{Z}_{2}$ graded spaces, this
operator is a parity operator, i.e., $R_{L}^{2}={I}$, inducing a $\mathbb{Z}%
_{2}$ grading of the state space. This means that $\mathcal{H}$ splits as a
direct sum, $\mathcal{H}^{(0)}\oplus \mathcal{H}^{(1)}$, of two orthogonal
subspaces: the odd (even) sector $\mathcal{H}^{(0)}=\mathcal{C}$ ($\mathcal{H%
}^{(1)}$). This grading can be lifted to the operator algebra over $\mathcal{%
H}$ turning this (Lie) algebra into a super- or $\mathbb{Z}_{2}$-graded Lie
algebra. Operators commuting (anticommmuting) with $R_{L}$ are referred to
as even (odd). Let $X\in \mathrm{End}(\mathcal{H})$; the even sector of the
algebra is given by $\{X\;|\;[R_{L},X]=0\}$ (i.e., $R_{L}XR_{L}^{\dagger }=X$%
) and the odd sector of the algebra is given by $\{X\;|\;\{R_{L},X\}=0\}$
(i.e., $R_{L}XR_{L}^{\dagger }=-X$).

As an example, let us suppose that all $S_\gamma$ are in the odd sector of
the algebra. According to the pulse sequence, Eq.~(\ref{eq:pkick}), any $%
H_{SB} = S_\gamma\otimes B_\gamma$, odd, in the system-bath Hamiltonian will
be removed after a complete set of operations. Therefore when all $S_\gamma$
are odd, complete decoupling is achievable using only $R_L$.

Now consider a \emph{leakage-elimination operator} (LEO) as in \cite%
{Wu/etal:02} 
\begin{equation}
R_{L}=e^{i\phi }\left( 
\begin{array}{cc}
-I & 0 \\ 
0 & I%
\end{array}%
\right) ,  \label{eq:R_Lmat}
\end{equation}%
where the blocks have the same dimensions as in Eq.~(\ref{eq:EEL}) and $\exp
(i\phi )$ is an overall phase factor. This operator anticommutes with the
leakage operators $\{R_{L},L\}=0$, while $[R_{L},E]=[R_{L},E^{\bot }]=0$.
Clearly such a sequence exactly produces the grading of the algebra
described in the previous section. $R_{L}$ is an LEO since it follows that
the following (parity-kick) sequence eliminates the leakage errors: 
\begin{equation}
\lim_{n\rightarrow \infty }(e^{-iHt/n}R_{L}^{\dagger
}e^{-iHt/n}R_{L})^{n}=e^{-iH_{E}t}e^{-iH^{\bot }t},  \label{e1}
\end{equation}%
where $H_{E}$ ($H^{\perp }$) corresponds to part of the error algebra which
affects only $\mathcal{C}$ ($\mathcal{C}^{\perp }$). To physically implement
this, in practice one takes $n=1$ and makes the time $t$ very small compared
to the bath correlation time as discussed in Section \ref{sec:dd}. Eq.~(\ref%
{e1}) then holds to order $t^{2}$, and implies that one intersperses periods
of free evolution for time $t$ with $R_{L}$, $R_{L}^{\dagger }$ strong
pulses. The term $e^{-iH^{\bot }t}$ in Eq.~(\ref{e1}) has no effect on the
qubit subspace. The term $e^{-iH_{E}t}$ may result in logical errors, which
will have to be treated by other methods, e.g., concatenation with a QECC 
\cite{Preskill:97a,Knill:98,Lidar:PRL99}, or additional BB pulses \cite%
{Viola:99,Byrd/Lidar:01,Byrd/Lidar:ss}. Therefore, in order to eliminate 
\emph{leakage}, we seek an LEO for a given encoding, which is obtainable
from a controllable system Hamiltonian $H_{S}$ acting for a time $\tau $,
i.e., $R_{L}=\exp (-iH_{S}\tau )$.

In \cite{Wu/etal:02}, several examples were given of physical systems which,
formally, have logical operations which are also naturally projective (i.e.,
they act as projections onto the code subspace). Such operations were termed
canonical. As mentioned above, in some situations, the physically available
(and controllable) interactions do not include operations which are also
projections. We will provide one important example in Section \ref{sec:4qdfs}%
, the four-qubit DFS in a solid state system which uses Heisenberg exchange
interactions for gating operations. In this case it is highly desirable to
have a more general method for producing the appropriate LEO which does not
include a projection. Next we will provide a generalized LEO which
circumvents the need for canonical gating operations, before discussing the
LEOs for the three- and four-qubit DFSs.


\subsection{Generalized LEO}

\label{sec:genLEO}

Generally, when a canonical logical operation is experimentally available,
we can construct an LEO using the methods from \cite{Wu/etal:02} where the
projector $\Pi _{\mathcal{C}}$ is redundant. In fact, an operator of the
following form serves as an LEO, 
\begin{equation}
R_{L}=\exp (-i\pi \sigma _{L}).
\end{equation}%
where $\sigma _{L}$ any operation such that $\sigma _{L}=\sigma
_{L}^{\dagger }$, $\sigma _{L}^{2}={I}$ on the code space, and $\sigma
_{L}\left\vert \psi \right\rangle =0$ for any $\left\vert \psi \right\rangle
\in \mathcal{C}^{\perp }$. The primary example is when $\sigma _{L}$ is a
canonical logical operation. Such is the case for the 3-qubit DFS which uses
Heisenberg exchange for logical gating operations, as will be shown later
(see also \cite{Wu/etal:02}).

A more general characterization of an LEO is the following. Let the
Hamiltonian for an LEO be given by 
\begin{equation}
H=\left( 
\begin{array}{cc}
H_{1} & 0 \\ 
0 & H_{2}%
\end{array}%
\right) ,
\end{equation}%
where $H_{1}$ acts on the code subspace and $H_{2}$ on the orthogonal
subspace. If $H_{1}$ is diagonal with even (odd) integers as the diagonal
elements and $H_{2}$ is diagonal with odd (even) integers as the diagonal
elements, then one may write the LEO as 
\begin{equation}
R_{L}=U\exp (-i\pi H)U^{\dagger }
\end{equation}%
where $U=U_{1}\oplus U_{2}$ is a direct sum (block diagonal). \textit{In
this case }$H$\textit{\ is not projective since it has non-zero eigenvalues
when acting on the subspace orthogonal to the code.} The effective LEO,
however, is unchanged, i.e., the form Eq.~(\ref{eq:R_Lmat}) is obtained,
which again produces a $\mathbb{Z}_{2}$ grading of the algebra and thus
eliminates leakage errors as desired. Such is the case for the four-qubit
DFS example in Section~\ref{sec:4qdfs}.


\subsection{Leakage Elimination to/from a Subspace}

Given the form of the operators that cause leakage, Eq.~(\ref{eq:EEL}) and
the form of the leakage elimination operator Eq.~(\ref{eq:R_Lmat}), we in
fact have the choice to eliminate leakage between $\mathcal{C}$ and $%
\mathcal{C}^{\perp }$ either by acting on $\mathcal{C}$, \emph{or} by action
on $\mathcal{C}^{\perp }$.

The advantage of the first, is that, in principle, we may use Theorem~\ref%
{th1} (\cite{Byrd/Lidar:ss,Byrd/Lidar:pqe02}), or the methods of \cite%
{Byrd/Lidar:ebb}, to eliminate all errors on the encoded state space, even
the logical errors. This requires (in many cases) acting on the codespace
with logical operations. If, however, we do not have the experimental
capabilities to implement the operations quickly enough for the given bath,
or if the operations are imperfect, such operations may make cause more
errors in the system \cite{Facchi/etal:04b}. However, we may choose to apply
the alternative of operating on $\mathcal{C}^{\perp }$. If the states are
properly confined to the codespace, then $\mathcal{C}^{\perp }$ states
should not be occupied, and decoupling pulses will have no effect. If the $%
\mathcal{C}^{\perp }$ states are becoming populated, then the decoupling
pulses applied to $\mathcal{C}^{\perp }$ will eliminate leakage. As long as
the decoupling pulses are properly constrained to act purely on $\mathcal{C}%
^{\perp }$ then this results in increased tolerance to other pulse
imperfections, in the sense that states in $\mathcal{C}$ are unaffected. As
we will see, in the case of the three- and four-qbuit DFSs, a large class of
errors are leakage type errors. Therefore decoupling with respect to $%
\mathcal{C}^{\perp }$ could serve as an effective error suppression method
in those cases where direct action on the codespace is inadvisable.


\section{LEOs AND THE 3-QUBIT DFS}

\label{sec:3qdfs}

The three qubit DFS encodes one logical qubit using a subsystem of three
physical qubits. It is the most efficient way in which to protect a single
logical qubit from collective errors of any type (bit-flip, phase-flip, or
both) \cite{Knill:00}. It has been shown that the exchange interaction is
sufficient for implementing a universal set of gating operations on this
system while preserving the DFS \cite{Kempe:00}. In this section we will
consider the effects of an LEO on this DFS and examine ways in which to
protect information which is encoded in this subsystem.


\subsection{3-qubit DFS}

\label{subsec:3qdfs}

We will represent the three-qubit DFS encoded qubit in the following way,
where $|0\rangle =|1/2\rangle ,|1\rangle =|-1/2\rangle $ are the two states
of a single spin-$1/2$ (Note that this convention is opposite to that of
reference \cite{Kempe:00}): 
\begin{equation}
\left( 
\begin{array}{c}
(\left\vert 010\right\rangle -\left\vert 100\right\rangle )/\sqrt{2} \\ 
(\left\vert 011\right\rangle -\left\vert 101\right\rangle )/\sqrt{2} \\ 
(2\left\vert 001\right\rangle -\left\vert 010\right\rangle -\left\vert
100\right\rangle )/\sqrt{6} \\ 
(-2\left\vert 110\right\rangle +\left\vert 011\right\rangle +\left\vert
101\right\rangle )/\sqrt{6} \\ 
\left\vert 000\right\rangle \\ 
(\left\vert 001\right\rangle +\left\vert 010\right\rangle +\left\vert
100\right\rangle )/\sqrt{3} \\ 
(\left\vert 011\right\rangle +\left\vert 101\right\rangle +\left\vert
110\right\rangle )/\sqrt{3} \\ 
\left\vert 111\right\rangle%
\end{array}%
\right) \overset{\mbox{\LARGE{\phantom{X}}}}{%
\begin{array}{c}
{\bigg\}}\left\vert 0_{L}\right\rangle \\ 
{\bigg\}}\left\vert 1_{L}\right\rangle \\ 
{\mbox{\LARGE{${\Bigg\}}$}}}\mathcal{C}^{\perp }%
\end{array}%
}  \label{eq:3DFS}
\end{equation}%
This notation means that $\left\vert 0_{L}\right\rangle =\alpha
_{0}(\left\vert 010\right\rangle -\left\vert 100\right\rangle )/\sqrt{2}%
+\beta _{0}(\left\vert 011\right\rangle -\left\vert 101\right\rangle )/\sqrt{%
2}$ (arbitrary superposition), and likewise $\left\vert 1_{L}\right\rangle
=\alpha _{1}(2\left\vert 001\right\rangle -\left\vert 010\right\rangle
-\left\vert 100\right\rangle )/\sqrt{6}+\beta _{1}(-2\left\vert
110\right\rangle +\left\vert 011\right\rangle +\left\vert 101\right\rangle )/%
\sqrt{6}$. The logical zero ($\left\vert 0_{L}\right\rangle $) and logical
one ($\left\vert 1_{L}\right\rangle $) comprise the code subspace $\mathcal{C%
}$. These states belong to the two $J=1/2$ irreducible representations
(irreps) of $SU(2)$. The coefficients are then the Wigner-Clebsch-Gordan
coefficients \cite{Bohm:qmbook} and the last 4 states comprise a $J=3/2$
representation of $SU(2)$. The two $J=1/2$ irreps can be distinguished by a
degeneracy label $\lambda =0,1$. Thus a basis state in the eight-dimensional
Hilbert space is fully identified by the three quantum numbers $|J,\lambda
,\mu \rangle $, where $\mu $ is the $z$-component of the total spin $J$. In
this notation we can write $\left\vert 0_{L}\right\rangle =\alpha
_{0}|1/2,0,1/2\rangle +\beta _{0}|1/2,0,-1/2\rangle $ and $\left\vert
1_{L}\right\rangle =\alpha _{1}|1/2,1,1/2\rangle +\beta
_{1}|1/2,1,-1/2\rangle $.


\subsection{Gating Operations for the 3-qubit DFS}

Physical gates were given in Refs. \cite{Kempe:00,DiVincenzo:00a,Hsieh:04}
and shown to be compatible with the DFS. The gates derived in Ref. \cite%
{Kempe:00} are generated by the Heisenberg exchange interaction between
pairs of physical qubits:%
\begin{equation}
E_{ij}=\frac{1}{2}(I+\vec{\sigma }_{i}\cdot \vec{\sigma }_{j}),
\end{equation}%
where $\vec{\sigma }=(\sigma _{x},\sigma _{y},\sigma _{z})$ is the vector of
Pauli matrices. As written $E_{ij}$ is the exchange operation between qubits 
$i$ and $j$, i.e., $E_{ij}|\phi \rangle _{i}|\psi \rangle _{j}=|\psi \rangle
_{i}|\phi \rangle _{j}$. The logical \textquotedblleft $X$%
\textquotedblright\ operation is given by \cite{Kempe:00} 
\begin{equation}
\bar{X}=\frac{1}{\sqrt{3}}(E_{23}-E_{13})=\left( 
\begin{array}{ccc}
0 & 1 & 0 \\ 
1 & 0 & 0 \\ 
0 & 0 & 0%
\end{array}%
\right) \otimes {I}_{2},
\end{equation}%
%
where we have labeled the rows and columns by the basis elements $%
\{|J=1/2,\lambda =0\rangle ,|J=1/2,\lambda =1\rangle ,|J=3/2,\lambda
=0\rangle \}$, and ${I}_{2}$ is the $2\times 2$ identity matrix which
accounts for the fact that the action of $E_{ij}$ is independent of the $\mu 
$ label ($z$-component of $J$) of the basis states. The logical
\textquotedblleft $Z$\textquotedblright\ operation is given by \cite%
{Kempe:00-fix} 
\begin{equation}
\bar{Z}=\frac{1}{3}(E_{13}+E_{23}-2E_{12})=\left( 
\begin{array}{ccc}
1 & 0 & 0 \\ 
0 & -1 & 0 \\ 
0 & 0 & 0%
\end{array}%
\right) \otimes {I}_{2},
\end{equation}%
and $\bar{Y}$ can be obtained from these two by commutation.

Recall that any operator can be decomposed in terms of a linear combination
of traceless, Hermitian matrices (plus the identity) with complex
coefficients. The exponentiation of the set of traceless, Hermitian
operators with real coefficients will give the set of unitary
transformations on the Hilbert space. Thus a Hermitian basis can be used to
decompose the error operators and Hamiltonians on the set of quantum states.

Here we will give a complete set of 64 operators on the space of three
qubits (63 traceless and the identity) and identify the logical operators,
collective operators, and leakage operators. This enables the identification
of various types of noise which can occur on the DFS and their effect on the
code. Our primary concern will be leakage errors and a leakage elimination
operator (LEO).

We note that it is also possible to eliminate all noises which are not
collective, thus producing the conditions for a DFS \cite%
{Zanardi:99a,Viola:00a,Wu/Lidar:cdfs,Byrd/Lidar:ebb}. The decomposition of
the errors in terms of basis elements and the identification of the types of
errors (leakage, collective, logical) on the code, will be useful for
identifying the type of error correction procedure which should be used to
correct errors affecting the code.


\subsection{The DFS-Basis}

As we saw above, collective errors on the DFS set of states act to mix the
two states within the $\left\vert 0_{L}\right\rangle $-subspace (and
simultaneously those within the $\left\vert 1_{L}\right\rangle $-subspace),
but do not mix the two subspaces with each other. Thus we use the labels $%
0_{L}$ and $1_{L}$ to identify the two-state subsystem used for storing
quantum information. This set of states is related to the standard set
(computational basis set, $|000\rangle ,...,|111\rangle $) through a
transformation we will denote $U_{dfs}$, and which can be read off directly
from Eq.~(\ref{eq:3DFS}): 
\begin{equation}
U_{df\!s}^{(3)}=\left( 
\begin{array}{cccccccc}
0 & 0 & \frac{1}{\sqrt{2}} & 0 & -\frac{1}{\sqrt{2}} & 0 & 0 & 0 \\ 
0 & 0 & 0 & \frac{1}{\sqrt{2}} & 0 & -\frac{1}{\sqrt{2}} & 0 & 0 \\ 
0 & \frac{2}{\sqrt{6}} & -\frac{1}{\sqrt{6}} & 0 & -\frac{1}{\sqrt{6}} & 0 & 
0 & 0 \\ 
0 & 0 & 0 & \frac{1}{\sqrt{6}} & 0 & \frac{1}{\sqrt{6}} & -\frac{2}{\sqrt{6}}
& 0 \\ 
1 & 0 & 0 & 0 & 0 & 0 & 0 & 0 \\ 
0 & \frac{1}{\sqrt{3}} & \frac{1}{\sqrt{3}} & 0 & \frac{1}{\sqrt{3}} & 0 & 0
& 0 \\ 
0 & 0 & 0 & \frac{1}{\sqrt{3}} & 0 & \frac{1}{\sqrt{3}} & \frac{1}{\sqrt{3}}
& 0 \\ 
0 & 0 & 0 & 0 & 0 & 0 & 0 & 1%
\end{array}%
\right) .
\end{equation}

Circuits for implementing logic gates on the $3$-qubit DFS were given in
Refs. \cite{Kempe:00,DiVincenzo:00a,Hsieh:04}. Here we take a different
approach. Rather than examining the DFS in the physical basis, as in Refs. 
\cite{Kempe:00,DiVincenzo:00a,Hsieh:04}, we examine the DFS in the DFS-basis
where the code and the effects of logical operators and leakage errors are
more transparent. This DFS-basis will be denoted with a tilde ($\tilde{%
\phantom{o}}$). This distinguishes a set of operators acting on the code
from the physical gating operators. The two sets of operators are related by
the DFS transformation $U_{dfs}\mathcal{O}U_{dfs}^{\dagger }=\tilde{\mathcal{%
O}}$. In the DFS-basis operators take the form%
\begin{equation}
\tilde{\mathcal{O}}=\left( 
\begin{tabular}{ll|l}
$\tilde{\mathcal{O}}_{0,0}^{1/2,1/2}$ & \multicolumn{1}{|l|}{$\tilde{%
\mathcal{O}}_{0,1}^{1/2,1/2}$} & $\tilde{\mathcal{O}}_{0,0}^{1/2,3/2}$ \\ 
\cline{1-2}
$\tilde{\mathcal{O}}_{1,0}^{1/2,1/2}$ & \multicolumn{1}{|l|}{$\tilde{%
\mathcal{O}}_{1,1}^{1/2,1/2}$} & $\tilde{\mathcal{O}}_{1,0}^{1/2,3/2}$ \\ 
\hline
$\tilde{\mathcal{O}}_{0,0}^{3/2,1/2}$ & $\tilde{\mathcal{O}}_{0,1}^{3/2,1/2}$
& $\tilde{\mathcal{O}}_{0,0}^{3/2,3/2}$%
\end{tabular}%
\right) ,  \label{eq:Op-DFS}
\end{equation}%
where $\tilde{\mathcal{O}}_{\lambda _{1},\lambda _{2}}^{J_{1},J_{2}}$ is a $%
(2J_{1}+1)\times (2J_{2}+1)$-dimensional block coupling $|J_{1},\lambda
_{1}\rangle $ states with $|J_{2},\lambda _{2}\rangle $ states. In the
DFS-basis, states are represented, using the notation of subsection~\ref%
{subsec:3qdfs}, as vectors of the form $(\alpha _{0},\beta _{0},\alpha
_{1},\beta _{1},\gamma _{-3/2},\gamma _{-1/2},\gamma _{1/2},\gamma
_{3/2})^{t}$, where the $\alpha ,\beta $-coefficients belong to the $J=1/2$
subsystems, and where the $\gamma $-coefficients belong to the $J=3/2$
subspace.

The advantage of the DFS-basis is this: In the DFS-basis every operator can
be decomposed into a tensor product of the form $\tilde{\mathcal{O}}_{\mu
}\otimes \tilde{\mathcal{O}}_{\lambda }\otimes \tilde{\mathcal{O}}_{J}$,
where each operator acts on the corresponding quantum number in a state $%
|J,\lambda ,\mu \rangle $.


\subsection{Decomposition of the Algebra}

\label{sec:basis3q}

In \cite{Wu/etal:02}, leakage errors between blocks $B$ and $C$ were
treated. Here we will carry this analysis further and investigate the types
of errors which may arise according to the algebraic decomposition and their
effect on the code. Note that the matrices in Eq.~(\ref{eq:EEL}) are $%
8\times 8$ matrices with $B,C,D,F$ all being $4\times 4$ blocks. Out of
these, only $D$ and $F$ represent leakage processes, so that there are a
total of $32$ independent such errors. In the DFS-basis these leakage errors
between the $J=1/2$ and $J=3/2$ subspaces have a simple representation. In
terms of Eq.~(\ref{eq:Op-DFS}) they appear as%
\begin{equation}
\tilde{\mathcal{O}}_{\mathrm{leak}}=\left( 
\begin{tabular}{ll|l}
& \multicolumn{1}{|l|}{} & $\tilde{\mathcal{O}}_{0,0}^{1/2,3/2}$ \\ 
\cline{1-2}
& \multicolumn{1}{|l|}{} & $\tilde{\mathcal{O}}_{1,0}^{1/2,3/2}$ \\ \hline
$\tilde{\mathcal{O}}_{0,0}^{3/2,1/2}$ & $\tilde{\mathcal{O}}_{0,1}^{3/2,1/2}$
& 
\end{tabular}%
\right) .  \label{eq:Op-DFS-leak}
\end{equation}%
The non-zero, off-diagonal blocks are $2\times (2(1/2)+1)\times
(2(3/2)+1)=4\times 4$ matrices, while $\tilde{\mathcal{O}}_{\mathrm{leak}}$
is $8\times 8$. It is then clear that we can construct an operator basis for
the leakage errors using the DFS-basis as follows: 
\begin{equation}
\tilde{X}\otimes \tilde{\mathcal{O}}_{\lambda }\otimes \tilde{\mathcal{O}}%
_{J},\;\;\;\;\mathrm{or}\;\;\;\;\tilde{Y}\otimes \tilde{\mathcal{O}}%
_{\lambda }\otimes \tilde{\mathcal{O}}_{J},  \label{eq:leaks}
\end{equation}%
where $\tilde{\mathcal{O}}_{\lambda },\tilde{\mathcal{O}}_{J}\in \left\{ 
\tilde{{I}},\tilde{X},\tilde{Y},\tilde{Z}\right\} $, i.e., each $\tilde{%
\mathcal{O}}_{i}$,$\;i=\lambda ,\mu $ is a Pauli matrix, or the identity
matrix, in the DFS-basis. The role of $\tilde{X}$ and $\tilde{Y}$ (which act
on the $J$ factor) is to put the $4\times 4$ matrix $\tilde{\mathcal{O}}%
_{\lambda }\otimes \tilde{\mathcal{O}}_{J}$ on the off-diagonal, as in Eq.~(%
\ref{eq:Op-DFS-leak}).

Similarly, a logical operator takes the form%
\begin{equation}
\tilde{\mathcal{O}}_{\mathrm{logic}}=\left( 
\begin{tabular}{ll|l}
$\tilde{\mathcal{O}}_{0,0}^{1/2,1/2}$ & \multicolumn{1}{|l|}{$\tilde{%
\mathcal{O}}_{0,1}^{1/2,1/2}$} &  \\ \cline{1-2}
$\tilde{\mathcal{O}}_{1,0}^{1/2,1/2}$ & \multicolumn{1}{|l|}{$\tilde{%
\mathcal{O}}_{1,1}^{1/2,1/2}$} &  \\ \hline
&  & 
\end{tabular}%
\right) ,
\end{equation}%
(a non-zero $4\times 4$ block) and, e.g., the logical $\sigma _{x}$ appears
as%
\begin{equation}
\bar{\sigma}_{x}=\left( 
\begin{tabular}{ll|l}
& \multicolumn{1}{|l|}{$I$} &  \\ \cline{1-2}
$I$ & \multicolumn{1}{|l|}{} &  \\ \hline
&  & 
\end{tabular}%
\right) .
\end{equation}%
Thus, the logical basis elements for the 3-qubit DFS code are represented
simply by 
\begin{gather}
\bar{\sigma}_{x}=\frac{1}{2}(\tilde{{I}}+\tilde{Z})\otimes \tilde{X}\otimes 
\tilde{{I}}  \notag \\
\bar{\sigma}_{y}=\frac{1}{2}(\tilde{{I}}+\tilde{Z})\otimes \tilde{Y}\otimes 
\tilde{{I}}  \notag \\
\bar{\sigma}_{z}=\frac{1}{2}(\tilde{{I}}+\tilde{Z})\otimes \tilde{Z}\otimes 
\tilde{{I}}.  \label{eq:logicalops}
\end{gather}%
The factor $(\tilde{{I}}+\tilde{Z})/2$ in these tensor products acts as a
projection onto the codespace. Therefore these operations act as ordinary
Pauli matrices on $\mathcal{C}$, or the $0_{L}-1_{L}$ block of the code and
are canonical operators in our sense (note that they also preserve the $\mu $
factors of $|0_{L}\rangle ,|1_{L}\rangle $). As we have seen in the previous
section, these can be implemented using Heisenberg exchange interactions
between qubits. Thus \textit{a canonical LEO is experimentally available in
systems which use Heisenberg exchange operations for gating and are encoded
in the 3-qubit DFS}.

The DFS logical states $\left\vert \Psi _{L}\right\rangle =a\left\vert
0_{L}\right\rangle +b\left\vert 1_{L}\right\rangle $ are, by construction,
invariant under collective errors. The (unnormalized) generators of
collective errors are $S_{\alpha }=\sum \sigma _{i}^{\alpha }$, $\alpha
=x,y,z$. To express these operators in the DFS basis, we transform by $%
U_{df\!s}$: 
\begin{gather}
S_{X}=(\tilde{{I}}+\tilde{Z})\tilde{{I}}\tilde{X}+\sqrt{3}(\tilde{{I}}-%
\tilde{Z})\tilde{{I}}\tilde{X}\;\;\;\;\;\;\;\;\;\;\;\;\;\;\;\;  \notag \\
+(\tilde{{I}}-\tilde{Z})\tilde{X}\tilde{X}+(\tilde{{I}}-\tilde{Z})\tilde{Y}%
\tilde{Y}\;\;\;\;\;\;  \label{eq:collerrorsX} \\
S_{Y}=(\tilde{{I}}+\tilde{Z})\tilde{{I}}\tilde{Y}+\sqrt{3}(\tilde{{I}}-%
\tilde{Z})\tilde{{I}}\tilde{Y}\;\;\;\;\;\;\;\;\;\;\;\;\;\;\;\;  \notag \\
+(\tilde{{I}}-\tilde{Z})\tilde{X}\tilde{Y}+(\tilde{{I}}-\tilde{Z})\tilde{Y}%
\tilde{X}\;\;\;\;\;\;  \label{eq:collerrorsY} \\
S_{Z}=(\tilde{{I}}+\tilde{Z})\tilde{{I}}\tilde{Z}+(\tilde{{I}}-\tilde{Z})%
\tilde{{I}}\tilde{Z}+(\tilde{{I}}-\tilde{Z})\tilde{Z}\tilde{{I}}  \notag \\
=\tilde{{I}}\tilde{{I}}\tilde{Z}+(\tilde{{I}}-\tilde{Z})\tilde{Z}\tilde{{I}}%
.\;\;\;\;\;\;\;\;\;\;\;\;\;\;\;\;\;\;\;\;\;\;\;\;  \label{eq:collerrorsZ}
\end{gather}%
This may appear to be a less convenient form than the form of the operators
in the physical basis, but in fact the interpretation is quite simple. For
example, consider the term $(\tilde{{I}}+\tilde{Z})\tilde{{I}}\tilde{X}$
appearing in $S_{X}$; in the DFS\ basis it is represented by 
\begin{equation}
\left( 
\begin{tabular}{ll|l}
$X$ & \multicolumn{1}{|l|}{} &  \\ \cline{1-2}
& \multicolumn{1}{|l|}{$X$} &  \\ \hline
&  & 
\end{tabular}%
\right),
\end{equation}%
i.e., it corresponds to an identical action on the two $J=1/2$ subsystems,
which is the signature of a collective error. Of course, the terms $(\tilde{{%
I}}+\tilde{Z})\tilde{{I}}\tilde{Y},(\tilde{{I}}+\tilde{Z})\tilde{{I}}\tilde{Z%
}$ have a similar interpretation. The remaining terms appearing in Eqs.~(\ref%
{eq:collerrorsX})-(\ref{eq:collerrorsZ}) are all projections on the
orthogonal subspace $\mathcal{C}^{\perp}$, since they contain the factor $(%
\tilde{{I}}-\tilde{Z})$. Thus they do not cause any errors on the DFS.

These observations facilitate the completion of the basis. Consider next the
basis for operations on the orthogonal subspace $\mathcal{C}^{\perp}$ which
are analogous to the logical operations on $\mathcal{C}$. These can be
determined from the set (\ref{eq:logicalops}) by simply replacing the
projector $(\tilde{{I}}+\tilde{Z})$ by $(\tilde{{I}}-\tilde{Z})$. However,
in order to keep all basis elements linearly independent, in particular
orthogonal to $S_{Z}$, we must modify $\bar{\sigma}_{z}^{\perp }$: 
\begin{gather}
\bar{\sigma}_{x}^{\perp }=(\tilde{{I}}-\tilde{Z})\tilde{X}\tilde{{I}}  \notag
\\
\bar{\sigma}_{y}^{\perp }=(\tilde{{I}}-\tilde{Z})\tilde{Y}\tilde{{I}}  \notag
\\
\bar{\sigma}_{z}^{\perp }=2\tilde{{I}}\tilde{{I}}\tilde{Z}-(\tilde{{I}}-%
\tilde{Z})\tilde{Z}\tilde{{I}}.  \label{eq:cperperrs}
\end{gather}

We now complete the 64 element orthogonal basis for the algebra \cite%
{MLEOs:comment2} Consider first the collective $X$ error, Eq.~(\ref%
{eq:collerrorsX}). The following four elements, appearing in $S_{X}$, and
whose interpretation as errors that either leave the DFS invariant or
annihilate it was discussed above, span a 4-dimensional subspace of the
algebra: 
\begin{equation}
(\tilde{{I}}+\tilde{Z})\tilde{{I}}\tilde{X},(\tilde{{I}}-\tilde{Z})\tilde{{I}%
}\tilde{X},(\tilde{{I}}-\tilde{Z})\tilde{X}\tilde{X},(\tilde{{I}}-\tilde{Z})%
\tilde{Y}\tilde{Y}  \label{eq:basiscollerrors}
\end{equation}%
To find a set of operators which are mutually orthogonal and orthogonal to $%
S_{X}$, we use the following procedure. First we normalize the collective $X$
error (with an overall factor of $1/\sqrt{6}$). We then require that the set
of mutually orthogonal operators,({\ref{eq:basiscollerrors}}), be taken from
the set of orthogonal baisis elements, to another set of orthogonal
elements. The appropriate mapping is an element of $SO(4)$ since it maps
four orthonormal vectors to four orthonormal vectors. Therefore, form an $%
SO(4)$ matrix whose first column elements are all $1/\sqrt{6}$, and whose
remaining three columns provide the following coefficients (the set formed
in this way is not unique): 
\begin{eqnarray}
S_{X_{1}} &=&\frac{1}{\sqrt{30}}(\tilde{{I}}+\tilde{Z})\tilde{{I}}\tilde{X}+%
\frac{1}{\sqrt{10}}(\tilde{{I}}-\tilde{Z})\tilde{{I}}\tilde{X}  \notag \\
&&+\frac{1}{\sqrt{30}}(\tilde{{I}}-\tilde{Z})\tilde{X}\tilde{X}-\sqrt{\frac{5%
}{6}}\;(\tilde{{I}}-\tilde{Z})\tilde{Y}\tilde{Y} \\
S_{{X}_{2}} &=&-\frac{\sqrt{3}}{2}\;(\tilde{{I}}+\tilde{Z})\tilde{{I}}\tilde{%
X}+\frac{1}{2}(\tilde{{I}}-\tilde{Z})\tilde{{I}}\tilde{X} \\
S_{X_{3}} &=&-\frac{1}{2\sqrt{5}}(\tilde{{I}}+\tilde{Z})\tilde{{I}}\tilde{X}-%
\frac{1}{2}\sqrt{\frac{3}{5}}\;(\tilde{{I}}-\tilde{Z})\tilde{{I}}\tilde{X} 
\notag \\
&&+\frac{2}{\sqrt{5}}\;(\tilde{{I}}-\tilde{Z})\tilde{X}\tilde{X},
\end{eqnarray}%
The result is three additional orthonormal basis elements which all act
trivially on the DFS.

The same procedure can be used for the collective $Y$ error [Eq.~(\ref%
{eq:collerrorsY})] to obtain the following set of trivially-acting errors: 
\begin{eqnarray}
S_{Y_{1}} &=&\frac{1}{\sqrt{30}}(\tilde{{I}}+\tilde{Z})\tilde{{I}}\tilde{Y}+%
\frac{1}{\sqrt{10}}(\tilde{{I}}-\tilde{Z})\tilde{{I}}\tilde{Y}  \notag \\
&&+\frac{1}{\sqrt{30}}(\tilde{{I}}-\tilde{Z})\tilde{X}\tilde{Y}-\sqrt{\frac{5%
}{6}}\;(\tilde{{I}}+\tilde{Z})\tilde{Y}\tilde{X} \\
S_{Y_{2}} &=&-\frac{\sqrt{3}}{2}\;(\tilde{{I}}+\tilde{Z})\tilde{{I}}\tilde{Y}%
+\frac{1}{2}(\tilde{{I}}-\tilde{Z})\tilde{{I}}\tilde{Y} \\
S_{Y_{3}} &=&-\frac{1}{2\sqrt{5}}(\tilde{{I}}+\tilde{Z})\tilde{{I}}\tilde{Y}-%
\frac{1}{2}\sqrt{\frac{3}{5}}\;(\tilde{{I}}-\tilde{Z})\tilde{{I}}\tilde{Y} 
\notag \\
&&+\frac{2}{\sqrt{5}}\;(\tilde{{I}}-\tilde{Z})\tilde{X}\tilde{Y}.
\end{eqnarray}%
Some of the lack of symmetry can be remedied by a different choice for the $%
SO(4)$ matrix. However, no significant simplification is obtained. The
(diagonal) collective $Z$ error may be completed by finding the remaining
elements in the set of eight diagonal elements which span the subspace of
diagonal elements of the algebra (the Cartan subalgebra). We now choose
elements to complete the basis, including this set. We start with the
following diagonal matrices 
\begin{equation}
\tilde{Z}\tilde{{I}}\tilde{Z},\;\;\tilde{Z}\tilde{{I}}\tilde{{I}},
\label{eq:diags}
\end{equation}%
which clearly also act as collective errors. Counting indicates there are $%
14 $ remaining basis elements. These can be taken to be 
\begin{gather}
(\tilde{{I}}-\tilde{Z})\tilde{X}\tilde{Z},\;\;(\tilde{{I}}-\tilde{Z})\tilde{Y%
}\tilde{Z},\;\;(\tilde{{I}}-\tilde{Z})\tilde{Z}\tilde{X},  \notag \\
(\tilde{{I}}-\tilde{Z})\tilde{Z}\tilde{Y},\;\;(\tilde{{I}}-\tilde{Z})\tilde{Z%
}\tilde{Z},  \label{eq:cperperrors}
\end{gather}%
which act to annihilate the DFS; and 
\begin{gather}
(\tilde{{I}}+\tilde{Z})\tilde{X}\tilde{X},\;\;(\tilde{{I}}+\tilde{Z})\tilde{X%
}\tilde{Y},\;\;(\tilde{{I}}+\tilde{Z})\tilde{X}\tilde{Z},  \notag \\
(\tilde{{I}}+\tilde{Z})\tilde{Y}\tilde{X},\;\;(\tilde{{I}}+\tilde{Z})\tilde{Y%
}\tilde{Y},\;\;(\tilde{{I}}+\tilde{Z})\tilde{Y}\tilde{Z},  \notag \\
(\tilde{{I}}+\tilde{Z})\tilde{Z}\tilde{X},\;\;(\tilde{{I}}+\tilde{Z})\tilde{Z%
}\tilde{Y},\;\;(\tilde{{I}}+\tilde{Z})\tilde{Z}\tilde{Z},  \label{eq:Cerrors}
\end{gather}%
which act non-trivially on the DFS in that they mix the $\mu $ factors of $%
|0_{L}\rangle ,|1_{L}\rangle $.

One may verify that we have enumerated $64$ basis elements, with the
property that they are trace orthogonal and therefore span the space of
three-qubit operators. These were chosen compatible with the appropriate
subspaces of the three-qubit DFS. The advantage of explicitly listing a
complete set of basis elements is that we may now classify the operators and
their actions on the code space.


\subsection{Classification of Errors}

In this subsection we classify the different types of operators in terms of
how they affect the three-qubit DFS code. We will then discuss the effect of
asymmetric exchange operations on this code in the next subsection.

As stated previously \cite{MLEOs:comment1}, we will not consider the errors
of type (\ref{eq:cperperrors}) or (\ref{eq:cperperrs}) as elements of the
algebra which would cause irrevocable loss of information (note, however,
that one could in principle remove them by symmetrizing with respect to $%
\mathcal{C}^{\perp}$). In addition, we have previously classified basis
elements Eq.~(\ref{eq:logicalops}) as logical operations which give rise to
logical errors when they are contained in the error algebra. The elements of
the form Eq.~(\ref{eq:leaks}) are a basis for the leakage type errors, while
terms of the form Eq.~(\ref{eq:collerrorsX}), (\ref{eq:collerrorsY}) and (%
\ref{eq:collerrorsZ}) are collective operations and will not affect the DFS
encoded states. The two diagonal elements (\ref{eq:diags}) do not cause
leakage and do not add anything beyond logical type errors. The remaining
operations need to be interpreted.

After some consideration, it is clear that the $S_{X_{i}},S_{Y_{i}}$
operators are actually elements of the stabilizer subgroup. For a DFS this
is defined as \cite{Kempe:00} 
\begin{equation}
\hat{D}(\vec{v})=\exp \left[ \sum_{\alpha }v_{\alpha }(\hat{S}_{\alpha
}-I\otimes M_{\alpha })\right]
\end{equation}%
where the elements of the vector $\vec{v}$, $v_{\alpha }$ are complex
numbers, the $\hat{S}_{\alpha }$ come from the set which generates the
collective error algebra $\mathcal{A}$ (we can include $H_{S}$ as $S_{0}$)
and $M_{\alpha }$ is a matrix which only mixes non-code indices. Thus the
linear combinations that are used in the sum in the argument of the
exponential will span the space of the primitives in the collective errors
and the $S_{X_{i}},S_{Y_{i}}$ are included in these linear combinations.
Another way in which to see that these operators do not affect the code
space is by acting with them on the code. The terms in $S_{X_{i}},$ and $%
S_{Y_{i}}$ of the form $(I+\tilde{Z})\ast \ast $ are the only ones which act
on the code space (with no effect), the others act on the orthogonal
complement $\mathcal{C}^{\perp }$. Thus these operations do not affect the
code and are thus elements of the stabilizer.

The remaining errors (\ref{eq:Cerrors}) are products of elements of the
logical operations and the collective operations. In this case $g \sim 
\mathcal{A}\mathcal{A}^\prime$ which is a product of an operator which does
nothing to the code (a stabilizer element) 
and a logical error.

Now that the complete set of basis elements spanning the three-qubit DFS has
been explicitly represented, and we have identified the action of these
operators on the code, we may discuss their significance in a more practical
setting. We will next decompose the errors that are seen as obstacles to
building practical solid state quantum computing systems and discuss the
affect of these errors in terms of the complete set of operations on the
code space.


\subsection{Errors in Quantum Dot Qubits}

A dominant type of error in solid-state quantum dot quantum computing
architectures arises from spin-orbit interactions \cite{Kavokin:01}.
Spin-orbit interactions couple charge degrees of freedom associated with the
orbital wave functions to the spin degrees of freedom used to store and
manipulate information. Since charge often interacts much more strongly with
the environment \cite{Burkard:99}, spin-orbit interactions give rise to
decoherence in these devices.

There are several ways in which to treat the gating errors which arise due
to anisotropic exchange interactions. One way is to treat them with a QECC,
which, as stated in the introduction, requires a substantial qubit overhead.
A second way is to use shaped pulses \cite{Tian:00}. A third way is to use
dressed qubits \cite{Wu/Lidar:03b}. The spin-orbit interaction can also be
used to construct a universal gate set \cite%
{Wu/Lidar:02c,Stepanenko/Bonesteel}. Here we assume the form of spin-orbit
interactions which give rise to errors which are of the same form as the
asymmetric, or anisotropic exchange. However, we treat these as decohering
(causing information loss), rather than unitary errors.

Consider a bilinear coupling in the physical basis of the form 
\begin{equation}
H_{SB}^{(2)}=\sum_{i<j}\sum_{\alpha ,\beta =\{x,y,z\}}g_{ij}^{\alpha \beta
}\sigma _{i}^{\alpha }\sigma _{j}^{\beta }\otimes B_{ij}^{\alpha \beta },
\end{equation}%
where $g_{ij}^{\alpha \beta }$ is a rank-$2$ tensor. The symmetrization
procedure of \cite{Wu/Lidar:cdfs}, that prepares collective decoherence
conditions, applies only to linear coupling, so will not work in this case.
In this case we must consider the possibility of leakage. The bilinear term $%
g_{ij}^{\alpha \beta }\sigma _{i}^{\alpha }\sigma _{j}^{\beta }$ can be
decomposed into (i) a scalar $g\vec{\sigma}_{i}\cdot \vec{\sigma}_{j}$,
which is proportional to the Heisenberg exchange operator and thus has the
effect of logical errors $E$; (ii) a rank-$1$ tensor 
\begin{equation}
\vec{\beta}\cdot (\vec{\sigma}_{i}\times \vec{\sigma}_{j});
\label{eq:sserr1}
\end{equation}%
(iii) a rank-$1$ tensor 
\begin{equation}
\left( \vec{\sigma}_{i}\cdot \vec{\gamma}_{i}\right) \left( \vec{\sigma}%
_{j}\cdot \vec{\gamma}_{j}\right)  \label{eq:sserr2}
\end{equation}%
which cannot couple the two $S=1/2$ states to each other, but can couple
them to $S=3/2$ states causing leakage. Thus we see that the $S=3/2$
subspace acts as a source for leakage [from (ii)\ and (iii)], and that there
is also the possibility of (non-collective) errors [from (ii)] which do not
have the same effect on the $|0_{L}\rangle ,|1_{L}\rangle $ states and
therefore cause logical errors. Clearly, higher-order interactions $%
H_{SB}^{(n)}$ with $n>2$ can cause similar leakage and logical errors.

We will now examine the errors (\ref{eq:sserr1}) and (\ref{eq:sserr2}) in
the basis for encoded operations and thus identify their effect on the code
space. To be specific, the errors in Eqs.~(\ref{eq:sserr1}) and (\ref%
{eq:sserr2}) will be transformed into the tilde basis, which acts on the
code space, by $U_{df\!s}E_{s}U_{df\!s}^{\dagger }$, where $E_{s}$ is either
of the form (\ref{eq:sserr1}) or (\ref{eq:sserr2}). We will then decompose
these operations in the DFS (tilde) basis of Section \ref{sec:basis3q}. Let
us first consider errors of the form (\ref{eq:sserr1}). We will consider
only at the errors on physical qubits one and two and neglect errors of the
type (\ref{eq:leaks}). After omitting these two types of errors, the
remaining errors are of the form 
\begin{equation}
\frac{1}{\sqrt{3}}[\beta _{12}^{x}(I+\tilde{Z})\tilde{Y}\tilde{X}+\beta
_{12}^{y}(I+\tilde{Z})\tilde{Y}\tilde{Y}+\beta _{12}^{z}({I}+\tilde{Z})%
\tilde{Y}\tilde{Z}].  \label{eq:crosserrs-leakage}
\end{equation}
Assuming that we can remove leakage errors using an appropriate LEO, our
goal is to analyze the remaining errors to see how they affect the encoded
qubits.

Note first that in Eq.~(\ref{eq:crosserrs-leakage}) we are examining errors
caused by interactions between qubits one and two. If only errors between
qubits one and two are present, there exists an inherent asymmetry in the
noise, thus we would not expect a DFS to protect against this type of error.
However, it turns out that the basis elements (primitives) here are the same
as those found in the interactions between qubits two and three and in
qubits one and three. Cancellation or partial cancellation of this type of
error will occur when the interactions between pairs of qubits have combine
to form only stabilizer operations, a situation which could be achieved
through material/environment engineering or by dynamical decoupling
symmetrization with respect to the code space. Here we will only consider
the asymmetric case of a pair of qubits (qubits 1 and 2).

How should we interpret the errors in Eq.~(\ref{eq:crosserrs-leakage})?
Consider the first term. Examining the logical errors, we see that the error
present in Eq.~(\ref{eq:crosserrs-leakage}) is proportional to $\bar{\sigma}%
_{y}\times S_{X}$ less those terms which are of the form $(I-\tilde{Z})\ast
\ast $. Thus it is a logical $Y$ multiplied by a collective $X$ error less
those errors which act only on $C^{\perp }$. The other terms are also
proportional to a logical $Y$ operation. This indicates that we may protect
against all of the remaining errors with the concatenation of this DFS by a
QECC which protects against logical $Y$ errors. This can be accomplished by
using a 3-to-1 QECC encoding. This may well be a significant advantage over
a pure QECC given available resources in solid-state implementations of
quantum computing.

Now we examine errors of the form (\ref{eq:sserr2}). Here, as before, we
will analyze only the affect of this type of error on physical qubits one
and two. We will also again omit the leakage errors which are of the form (%
\ref{eq:leaks}). The remaining 10 terms of the DFS transformed errors (\ref%
{eq:sserr2}) are 
\begin{gather}
\frac{{\gamma _{1}^{x}}{\gamma _{2}^{x}}-{\gamma _{1}^{y}}{\gamma _{2}^{y}}}{%
2\sqrt{3}}(I-\tilde{Z})\tilde{X}I+\frac{{\gamma _{1}^{y}}{\gamma _{2}^{x}}+{%
\gamma _{1}^{x}}{\gamma _{2}^{y}}}{2\sqrt{3}}(I-\tilde{Z})\tilde{Y}I  \notag
\\
+\frac{(-{\gamma _{1}^{z}}{\gamma _{2}^{y}}+{\gamma _{1}^{y}}{\gamma _{2}^{z}%
})}{2\sqrt{3}}(I+\tilde{Z})\tilde{Y}\tilde{X}+\frac{{\gamma _{1}^{z}}{\gamma
_{2}^{x}}-{\gamma _{1}^{x}}{\gamma _{2}^{z}}}{2\sqrt{3}}(I+\tilde{Z})\tilde{Y%
}\tilde{Y}  \notag \\
+\frac{-{\gamma _{1}^{y}}{\gamma _{2}^{x}}+{\gamma _{1}^{x}}{\gamma _{2}^{y}}%
}{2\sqrt{3}}(I+\tilde{Z})\tilde{Y}\tilde{Z}  \notag \\
+\frac{1}{3}(-{\gamma _{1}^{x}}{\gamma _{2}^{x}}-{\gamma _{1}^{y}}{\gamma
_{2}^{y}}-{\gamma _{1}^{z}}{\gamma _{2}^{z}})(I+\tilde{Z})\tilde{Z}I  \notag
\\
+\frac{{\gamma _{1}^{z}}{\gamma _{2}^{x}}+{\gamma _{1}^{x}}{\gamma _{2}^{z}}%
}{2\sqrt{3}}(I-\tilde{Z})\tilde{Z}\tilde{X}+\frac{({\gamma _{1}^{z}}{\gamma
_{2}^{y}}+{\gamma _{1}^{y}}{\gamma _{2}^{z}})}{2\sqrt{3}}(I-\tilde{Z})\tilde{%
Z}\tilde{Y}  \notag \\
+\frac{1}{6}(-{\gamma _{1}^{x}}{\gamma _{2}^{x}}-{\gamma _{1}^{y}}{\gamma
_{2}^{y}}+2{\gamma _{1}^{z}}{\gamma _{2}^{z}})(I-\tilde{Z})\tilde{Z}\tilde{Z}
\notag \\
+\frac{1}{3}(-\gamma _{1}^{x}\gamma _{2}^{x}-\gamma _{1}^{y}\gamma
_{2}^{y}-\gamma _{1}^{z}\gamma _{2}^{z})\tilde{Z}II.
\end{gather}%
%
Terms 1,2,7,8,9 act on $\mathcal{C}^{\perp } $ and thus have no affect on
the code. Term 10 is a combination of exchange operations which does not
affect the code subspace. Terms 3,4,5,6 act on the code space. They act
either as a logical error (term 6) or as a logical error composed with a
collective error. The logical errors must be treated using other methods. In
this case, we see that all errors other than term 6 give rise to logical $Y$
errors. To remove all errors in the system, we have different choices which
could be good alternatives depending upon the physical system. First, we
could eliminate the errors using more decoupling pulses. Second, we could
choose a system where the term 6 is negligible. This would enable complete
elimination of the errors with one other decoupling pulse. Third, if term 6
is negligible, we could treat the $Y$ errors with the concatenation of the
DFS with a QECC. The required QECC would use only three qubits to encode one.

Physical implementations will dictate whether decoupling pulses, materials
engineering, or a QECC is a valid option for the elimination of remaining
errors after leakage has been removed.


\section{LEOs AND THE 4-QUBIT DFS}

\label{sec:4qdfs}

The four qubit DFS encodes one logical qubit using a subspace of four
physical qubits. It protects a single logical qubit from collective errors
of any type (bit-flip, phase-flip, or both). It has been shown that the
exchange interaction is sufficient for implementing a universal set of
gating operations on this system while preserving the DFS \cite{Kempe:00}.
In this section we will answer the following questions: 1) Is there an LEO
available in solid-state implementations of qubits which relies on the
exchange interaction? 2) Is there a canonical set of such gates? 3) After
leakage is removed, can we completely remove all errors in asymmetric
exchange, and if so, how?


\subsection{The Four-qubit DFS}

The four-qubit DFS contains two singlet states for the representative qubit,
three triplets and a spin 2, or quintuplet. The singlet states, which
represent the logical zero and one of the DFS encoded qubit are given by 
\begin{equation}
S^0 =\left\vert 0_L\right\rangle= \frac{1}{2}(\left\vert 0101\right\rangle +
\left\vert 1010\right\rangle - \left\vert 0110\right\rangle - \left\vert
1001\right\rangle)
\end{equation}
and 
\begin{eqnarray}
S^1 &=& \left\vert 1_L\right\rangle  \notag \\
&=& \frac{\;1}{\sqrt{12}}(2\left\vert 0011\right\rangle + 2\left\vert
1100\right\rangle  \notag \\
& &- \left\vert 0110\right\rangle - \left\vert 1001\right\rangle -
\left\vert 0101\right\rangle - \left\vert 1010\right\rangle),
\end{eqnarray}
where as before, e.g., $\left\vert 0101\right\rangle = \left\vert
1/2\right\rangle\otimes \left\vert -1/2\right\rangle\otimes\left\vert
1/2\right\rangle\otimes\left\vert -1/2\right\rangle$ in the standard angular
momentum basis. $S^0$ and $S_1$ comprise the code space $\mathcal{C}$ for
this DFS and the remaining states are the states in $\mathcal{C}^\perp$.

The triplet states are given by 
\begin{eqnarray}
T_{11}^{1}=\left\vert 11\right\rangle &=&\frac{1}{2}(\left\vert
0100\right\rangle +\left\vert 1000\right\rangle -\left\vert
0001\right\rangle -\left\vert 0010\right\rangle )  \notag \\
T_{10}^{1}=\left\vert 10\right\rangle &=&\frac{1}{\sqrt{2}}(\left\vert
1100\right\rangle -\left\vert 0011\right\rangle ) \\
T_{1-1}^{1}=\left\vert 1-\!\!1\right\rangle &=&\frac{1}{2}(\left\vert
1110\right\rangle +\left\vert 1101\right\rangle -\left\vert
1011\right\rangle -\left\vert 0111\right\rangle )  \notag
\end{eqnarray}%
\begin{eqnarray}
T_{11}^{2} &=&\left\vert 11\right\rangle =\frac{1}{\sqrt{2}}(\left\vert
0001\right\rangle -\left\vert 0010\right\rangle )  \notag \\
T_{10}^{2} &=&\left\vert 10\right\rangle =\frac{1}{2}(\left\vert
1001\right\rangle +\left\vert 0101\right\rangle -\left\vert
1010\right\rangle -\left\vert 0110\right\rangle )  \notag \\
T_{1-1}^{2} &=&\left\vert 1-\!\!1\right\rangle =\frac{1}{\sqrt{2}}%
(\left\vert 1101\right\rangle -\left\vert 1110\right\rangle )
\end{eqnarray}%
\begin{eqnarray}
T_{11}^{3} &=&\left\vert 11\right\rangle =\frac{1}{\sqrt{2}}(\left\vert
0100\right\rangle -\left\vert 1000\right\rangle )  \notag \\
T_{10}^{3} &=&\left\vert 10\right\rangle =\frac{1}{2}(\left\vert
0110\right\rangle +\left\vert 0101\right\rangle -\left\vert
1010\right\rangle -\left\vert 1001\right\rangle )  \notag \\
T_{1-1}^{3} &=&\left\vert 1-\!\!1\right\rangle =\frac{1}{\sqrt{2}}%
(\left\vert 0111\right\rangle -\left\vert 1011\right\rangle )
\end{eqnarray}%
The spin-2 representation is given by the following set of states (a
quintuplet) 
\begin{eqnarray}
Q_{22} &=&\left\vert 22\right\rangle =\left\vert 0000\right\rangle  \notag \\
Q_{21} &=&\left\vert 10\right\rangle =\frac{1}{2}(\left\vert
1000\right\rangle +\left\vert 0100\right\rangle +\left\vert
0010\right\rangle +\left\vert 0001\right\rangle )  \notag \\
Q_{20} &=&\left\vert 1-\!\!1\right\rangle =\frac{\;1}{\sqrt{6}}(\left\vert
1100\right\rangle +\left\vert 1010\right\rangle  \notag \\
&&+\left\vert 1001\right\rangle +\left\vert 0110\right\rangle +\left\vert
0101\right\rangle +\left\vert 0011\right\rangle )  \notag \\
Q_{2-1} &=&\left\vert 2-\!\!1\right\rangle =\frac{1}{2}(\left\vert
0111\right\rangle +\left\vert 1011\right\rangle +\left\vert
1101\right\rangle +\left\vert 1110\right\rangle )  \notag \\
Q_{2-2} &=&\left\vert 2-2\right\rangle =\left\vert 1111\right\rangle
\end{eqnarray}

We will refer to this set of states as the set of DFS states with the
logical elements comprising $\mathcal{C}$ and the other states ($T$ and $Q$
states) comprising $\mathcal{C}^\perp$. In the latter part of this section
we will again use a DFS basis for the operators and a transformation $U_{dfs}
$ to change from the computational basis set of states and operators to the
DFS sets. Let us now discuss logical operations on the DFS.


\subsection{Gate Operations and LEOs}

Physical gates were given in Ref. \cite{Kempe:00} and shown to be compatible
with the DFS. These are given by the exchange interaction between pairs of
physical qubits. The logical \textquotedblleft $X$\textquotedblright\
operation is given by 
\begin{equation}
\bar{X}=\frac{1}{\sqrt{3}}(E_{23}-E_{13})
\end{equation}%
where, again, $E_{ij}$ is the exchange operation between qubits $i$ and $j$,
and $I_{2}$ is the $2\times 2$ identity matrix. The logical
\textquotedblleft $Z$\textquotedblright\ operation is given by 
\begin{equation}
\bar{Z}=-E_{12}
\end{equation}%
and $\bar{Y}$ can be obtained from these two. However, there is a distinct
difference between this set of logical operations and the analogous set in
Section \ref{sec:3qdfs}. The difference is that naturally occuring logical
operations on the three-qubit DFS have eigenvalue zero on the states in $%
\mathcal{C}^{\perp }$. This in not the case for the four-qubit DFS. For
example, all states in $\mathcal{C}^{\perp }$ have eigenvalues of $+1$ for
states in $\mathcal{C}^{\perp }$, with the exception of $T^{3}$ whose states
have eigenvalue $-1$ when acted on by $\bar{Z}$. These gates are not
\textquotedblleft canonical\textquotedblright\ in our sense and so the
projection onto the code subspace is not automatic. One must either use
another set of gating operations which are projective or use the generalized
LEO of Section \ref{sec:genLEO}. We explore each of these two possibilities
in the next two subsections.


\subsection{An LEO from Exchange}

As noted in \cite{Wu/etal:02} there exists physically available operations
which are ``canonical'' in many circumstances, meaning that they are
projective onto the code subspace. The definition of $\bar{Z}$ given in \cite%
{Kempe:00} (and the previous section) is not canonical. We therefore seek to
construct a physically available LEO. We do this using the definition of the
generalized LEO given in Section \ref{sec:genLEO}.

Let the (square of the) total spin angular momentum operator be denoted $%
\vec{S}^2$ with eigenvalue $S(S+1)$. Then 
\begin{equation*}
4\vec{S}^2 = \left(\sum_i \vec{\sigma}_i\right)^2,
\end{equation*}
where $\vec{\sigma}_i = (\sigma_i^x,\sigma_i^y,\sigma_i^z)$ are the Pauli
matrices acting on the $i^{\mbox{th}}$ qubit. Therefore 
\begin{equation*}
S^2/2 = \frac{12 + 2\sum_{i<j}\vec{\sigma}_i \cdot \vec{\sigma}_j}{8}
\end{equation*}
gives an appropriate LEO of the form given in Eq.~(\ref{eq:R_Lmat}). This
can be seen as follows. On the $S = 0$ (singlet) subspaces the operator
gives zero. On the $S=1$ subspaces the operator gives $1$ and on the $S=3$
subspace it gives $3$. Therefore the appropriate LEO is given by 
\begin{equation*}
R_L = \exp\{-i\pi S^2/2\},
\end{equation*}
which reproduces the LEO of Eq.~(\ref{eq:R_Lmat}).

Now consider a modified set of logical operations: 
\begin{gather}
\bar{X} \rightarrow \bar{X}^\prime = \bar{X} + S^2/2, \\
\bar{Y} \rightarrow \bar{Y}^\prime = \bar{Y} + S^2/2, \\
\bar{Z} \rightarrow \bar{Z}^\prime = \bar{Z} + S^2/2.
\end{gather}
This set may be used to obtain a appropriate LEO by exponentiation, e.g. 
\begin{equation*}
R_L = \exp\{-i\pi(\bar{Z}^\prime)\}.
\end{equation*}
Similarly we can construce an LEO using $\bar{X}^\prime$ or $\bar{Y}^\prime$%
. Since the operator $S2$ is composed of exchange interactions, it is also
experimentally available.


\subsection{A canonical LEO from exchange}

One may ask the question: what set of gates would be canonical if we use
only exchange operations? One way to answer this question is to do the
following calculation. Start with operations which act as $\bar{X}, \bar{Y}$
and $\bar{Z}$ and have eigenvalue zero on the states in $\mathcal{C}^\perp$.
These are operators in the DFS (i.e., the tilde) basis. Then use the DFS
transformation to transform from the DFS basis back to the physical basis to
find the set of physical interactions necessary to perform canonical gating
operations. We now use this procedure to find such an LEO.

The DFS basis for this system has logical operations which transform between
the two one-dimensional spin-0 representations. These may be represented by
ordinary Pauli matrices which act only on the $2\times 2$ block. The
operations which perform these logical operations are labelled using the
spin-0 index and a $0,1$ degeneracy index, $\mathcal{O}^{0,0}_{\lambda_1,%
\lambda,2}$, which in this case has only a single entry for each pair ${%
\lambda_1,\lambda,2}$. According to the definition, canonical logical
operations would have the following form in the DFS basis: 
\begin{equation}
\bar{\sigma}_{i}=\left( 
\begin{array}{cc}
\sigma_{i} & 0_{14\times 2} \\ 
0_{2\times 14} & 0_{14\times 14}%
\end{array}%
\right),
\end{equation}%
where $0_{m\times n}$ is an $m\times n$ matrix of zeros. This operation acts
simultaneously as a projector onto the code subspace and a Pauli operator on
the encoded state. Now, let the DFS transformation be given by $U_{df\!s}$,
the computational basis states be given by $\left\vert \psi
_{c}\right\rangle $, and the DFS states be given by $\left\vert \psi
_{df\!s}\right\rangle $: 
\begin{equation}
U_{df\!s}\left\vert \psi _{c}\right\rangle =\left\vert \psi
_{df\!s}\right\rangle .
\end{equation}%
Then the logical operations are related to the operations in the
computational basis by 
\begin{equation}
\bar{\sigma}_{i}\left\vert \psi _{df\!s}\right\rangle =U_{df\!s}\bar{\sigma}%
_{i}^{c}\left\vert \psi _{c}\right\rangle ,
\end{equation}%
where $\bar{\sigma}_{i}$ is the canonical logical operation in the logical
basis and $\bar{\sigma}_{i}^{c}$ is the logical operation in the
computational basis. This physical realization of the canonical operations
will be found using 
\begin{equation}
U_{df\!s}^{-1}\bar{\sigma}_{i}U_{df\!s}=\bar{\sigma}_{i}^{c}.
\end{equation}%
Using only the exchange operations between qubits $i$ and $j$, $E_{ij}=I+%
\vec{\sigma}_{i}\cdot \vec{\sigma _{j}}$, the canonical operations are given
by 
\begin{equation}
\bar{\sigma}_{x}^{c}=(2I-E_{13})(2I-E_{24})-(2I-E_{23})(2I-E_{14})
\end{equation}%
and 
\begin{eqnarray}
\bar{\sigma}_{z}^{c} &=&2(2I-E_{34})(2I-E_{12})-(2I-E_{13})(2I-E_{24}) 
\notag \\
&&-(2I-E_{23})(2I-E_{14})
\end{eqnarray}%
and the commutator of these two gives the third logical element. It is clear
from this form that 4-body interactions are required to construct canonical
logical operations using only exchange operations. Although there are
methods for constructing these from more fundamental interactions \cite%
{Wang/Zanardi:02,Wu/etal:BCS} so that they may be useful for some quantum
computing purposes (e.g. simulations) they are likely impratical for BB
controls due to time constraints. Alternatively, it has been shown that
these four-body interactions naturally arise in some systems with
significant effect and it has been suggested that one might take advantage
of these effects when designing gating operations \cite{Mizel/Lidar:04}.


\subsection{Errors in quantum dot qubits}

The detailed analysis of Section \ref{sec:3qdfs} becomes much more
complicated for the four-qubit DFS. The set of operators for the Hilbert
space of four qubits contains $(2^{4})^{2}=2^{8}=256$ basis elements. This
set obviously becomes prohibitively large for such a detailed analysis as
the number of qubits grows. However, there are several key observations
which summarize our three-qubit analysis which are quite general. (These can
also be seen from the algebraic decomposition of Section \ref{sec:alg}.) We
can see that the operations in the DFS basis may be classified as follows:
Logical operations, operations on $\mathcal{C}^{\perp }$, collective
operations, leakage operations, products of one or more of these types. As
in the treatment of the 3-qubit DFS, we will assume that the LEO can be
implemented so that leakage errors are irrelevant. In this case, we need
only remove logical operations (this will also remove operations which are
products of logical operations and another type). Therefore our goal is to
identify logical operations and then extract those operators in the
spin-orbit coupling terms which give rise to logical type errors, whether or
not they are combined with collective errors.

The extraction of logical operations and operations which are products of
logical operations with another are readily performed with {\small {\texttt{%
MATHEMATICA}}}. We note first that with an ordered logical, or DFS, basis
which contains the first two entries $\left\vert 0_{L}\right\rangle $ and $%
\left\vert 1_{L}\right\rangle $, will have logical basis elements for the
algebra of the form 
\begin{equation}
(I-\tilde{Z})(I-\tilde{Z})(I-\tilde{Z})\mathcal{O}_{i},
\end{equation}%
where $\mathcal{O}_{i}\in \{\tilde{X},\tilde{Y},\tilde{Z}\}$. This implies
that from Section~\ref{sec:LEOs}, $B$ is a $2\times 2$ matrix, $C$ is a $%
14\times 14$ matrix, $E$ is $2\times 14$ and $E^{\perp }$ is $14\times 2$.

Again we decompose the DFS transformed terms of the form Eqs.~(\ref%
{eq:sserr1}) and (\ref{eq:sserr2}) in terms of the DFS basis elements. In
this case the representations used for the DFS states are not completely
symmetric in the four qubits. We must therefore consider two pairs of
interactions, the interactions between qubits one and two, and between
qubits two and three.

The error arising from Eq.~(\ref{eq:sserr1}) between physical qubits 1 and
2, and neglecting leakage operations and operations on $\mathcal{C}^{\perp }$%
, consists of just one term: 
\begin{equation}
-\frac{1}{3\sqrt{2}}\left( \gamma _{1}^{x}\gamma _{2}^{x}+\gamma
_{1}^{y}\gamma _{2}^{y}+\gamma _{1}^{z}\gamma _{2}^{z}\right) (I+\tilde{Z}%
)(I+\tilde{Z})(I+\tilde{Z})\tilde{Z}.
\end{equation}%
There are two different terms for the Eq.~(\ref{eq:sserr2}) when acting
between qubits 2 and 3: 
\begin{gather}
\frac{1}{6\sqrt{2}}\left( \gamma _{1}^{x}\gamma _{2}^{x}+\gamma
_{1}^{y}\gamma _{2}^{y}+\gamma _{1}^{z}\gamma _{2}^{z}\right) (I+\tilde{Z}%
)(I+\tilde{Z})(I+\tilde{Z})\tilde{Z} \\
\frac{1}{2\sqrt{6}}\left( \gamma _{1}^{x}\gamma _{2}^{x}+\gamma
_{1}^{y}\gamma _{2}^{y}+\gamma _{1}^{z}\gamma _{2}^{z}\right) (I+\tilde{Z}%
)(I+\tilde{Z})(I+\tilde{Z})\tilde{X}.
\end{gather}

A rather stark difference between the three-qubit DFS and the four-qubit DFS
is the fact that no errors of the form Eq.~(\ref{eq:sserr2}) contribute to
logical errors on this code. We therefore need not correct this type of
error. These type of errors occur in zinc-blend type semiconductor
structures which have a broken inversion symmetry. Knowing that removing all
leakage errors can render this type of error insignificant is a important
advantage of this code.

Therefore, we again can correct all errors of the forms in Eqs.~(\ref%
{eq:sserr1}) and (\ref{eq:sserr2}) with the concatenation of a three-qubit
QECC or with one added decoupling pulse. However, we see that the
antisymmetric term, also known as a Dzyaloshinski-Moriya term, is not
present after the implementation of an appropriate LEO. Therefore, if this
term is a dominant source of errors in a particular implementation of
quantum dot quantum computing, then we recommend the four-qubit DFS over the
three qubit DFS.


\section{CONCLUSIONS}

We have given methods for the implementation of leakage elimination
operators LEOs which are, in many circumstances, physically available in
experiments. These LEOs eliminate a large and important class of errors;
those that would serve to destroy a subspace encoding. The methods for
producing these have shown promise in many experiments and here we have
generalized the method for producing such LEOs.

Symmetrization by dynamical decoupling can be performed by decoupling the
code from its orthogonal subspace or the orthogonal subspace from the code.
In the first case, we can, in principle, remove all errors from the code
space. While this is not true when the decoupling is performed with respect
to the orthogonal subspace, the advantage of this method is that it does not
introduce further errors if imperfect decoupling controls are used.

For the three- and four-qubit DFSs, we have given a method for producing
LEOs using the exchange interaction which is physically available in
solid-state qubit implementations of quantum computing. We have shown that
after leakage errors have been removed, the remaining bilinear couplings
which could give rise to errors may be treated with either a QECC or cycle
involving one extra decoupling pulse. In this case we can, in fact, produce
the conditions for a DFS even when no symmetry is present in the original
system. The four-qubit DFS has an advantage in this regard over the
three-qbuit DFS since eliminating leakage errors from the four-qubit DFS
removes the Dzyaloshinski-Moriya interaction errors which are present in
many semiconductors which have a broken inversion symmetry and are the main
part of the anisotropic exchange interaction.


\begin{acknowledgments}
D.A.L. acknowledges financial support from AFOSR (F49620-01-1-0468), the
Sloan Foundation, NSERC, and the Connaught Fund.
\end{acknowledgments}



\begin{thebibliography}{79}
\bibitem[{MLE({\natexlab{a}})}]{MLEOs:comment1}
\bibinfo{note}{In the case that we are discussing elements of the error algebra
  in the Kraus representation which are elements of ${\cal A}$, these actually
  correspond to projections onto the space orthogonal to projections onto
  $C^\perp$. However, we don't want to consider these here. We will consider
  their exponentials.}
\bibitem[{BBn()}]{BBnote}
\bibinfo{note}{{We should emphasize that although we are discussing a bang-bang
  pulse sequence, our arguments hold for any suitable set of decoupling
  pulses.}}
\bibitem[{Kem()}]{Kempe:00-fix}
\bibinfo{note}{{We note that the corresponding Eq.~(74) in Ref.~\cite{Kempe:00}
  contains incorrect numerical factors, which we have fixed here.}}
\bibitem[{MLE({\natexlab{b}})}]{MLEOs:comment2}
\bibinfo{note}{We are referring to trace orthogonality, i.e., using the
  Hilbert-Schmidt inner product.}
\end{thebibliography}


\end{document}